\documentclass[useAMS,usenatbib]{mn2e}
\usepackage{times}
\usepackage{url}
\usepackage{graphics,graphicx}
\usepackage{deluxetable}
\usepackage[usenames]{color}
\DeclareGraphicsExtensions{.pdf,.png,.jpg,.mps,.eps,.ps}
\usepackage{amsmath}
\usepackage{natbib}

\newcommand\suzaku{{\it Suzaku}}

\newcommand\xmm{{\it XMM-Newton}}

\newcommand\ks{{\rm~ks}}

\newcommand\mpc{{\rm~Mpc}}

\newcommand\kev{{\rm~keV}}
\newcommand\ev{{\rm~eV}}
\newcommand\kms{\ifmmode {\rm~km\ s}$^{-1}$ \else ~km s$^{-1}$\fi}
\newcommand\Hunit{\ifmmode {\rm~km\ s}$^{-1}$\ {\rm Mpc}$^{-1}$
        \else ~km s$^{-1}$ Mpc$^{-1}$\fi}
\newcommand\ctssec{\ifmmode {\rm~count\ s}$^{-1}$ \else ~count s$^{-1}$\fi}
\newcommand\ergsec{\ifmmode {\rm~erg\ s}$^{-1}$ \else
        ~erg s$^{-1}$\fi}
\newcommand\funit{\ifmmode {\rm~erg\ s}$^{-1}$\ ; {\rm cm}$^{-2}$ \else
        ~ergs s$^{-1}$ cm$^{-2}$\fi}
\newcommand\phflux{\ifmmode {\rm~photon\ s}$^{-1}$\  ; {\rm cm}$^{-2}$
        \else   ~photon s$^{-1}$ cm$^{-2}$\fi}
\newcommand\efluxA{\ifmmode {\rm~erg\ s}$^{-1}$\ ; {\rm cm}$^{-2}$\ ; {\rm
        \AA}$^{-1}$ \else ~erg s$^{-1}$ cm$^{-2}$ \AA$^{-1}$\fi}
\newcommand\efluxHz{\ifmmode {\rm~erg\ s}$^{-1}$\ ; {\rm cm}$^{-2}$\ ; {\rm
        Hz}$^{-1}$ \else ~erg s$^{-1}$ cm$^{-2}$ Hz$^{-1}$\fi}
\newcommand\cc{\ifmmode {\rm~cm}$^{-3}$ \else cm$^{-3}$\fi}
\newcommand\FWHM{\ifmmode {\rm~FWHM} \else ${\rm~FWHM}$\fi}
\newcommand\Msun{\ifmmode M_{\odot} \else $M_{\odot}$\fi}
\newcommand\Lsun{\ifmmode L_{\odot} \else $L_{\odot}$\fi}

\newcommand\hbeta{\ifmmode {\rm H}\beta \else H$\beta$\fi}
\newcommand\Kalpha{\ifmmode {\rm K}\alpha \else K$\alpha$\fi}
\newcommand\nh{\ifmmode N_{\rm H} \else N$_{\rm H}$\fi}

\makeatletter

\newcommand{\Rmnum}[1]{\expandafter\@slowromancap\romannumeral #1@}
\makeatother

\title {Variable X-ray reflection from 1H~0419--577}
\author [Pal \& Dewangan] {Main Pal\thanks { Email: mainpal@iucaa.ernet.in} and
Gulab \ C.\ Dewangan\thanks{Email: gulabd@iucaa.ernet.in} \\
Inter University Centre for Astronomy and Astrophysics (IUCAA), Pune 411007, India}

\begin{document}
\maketitle
\begin{abstract}
  We present detailed broadband X-ray spectral variability of a
  Seyfert 1 galaxy 1H~0419--577 based on an archival
  \suzaku{} observation in July 2007, a new \suzaku{} observation
  performed in January 2010 and the two latest \xmm{} observations
  from May 2010. All the observations show soft X-ray
  excess emission below $2\kev$ and both \suzaku{} observations show a
  hard X-ray excess emission above $10\kev$ when compared to a
  power-law. We have tested three physical models -- a complex partial
  covering absorption model, a blurred reflection model
      and an intrinsic disk Comptonization model. Among these three
      models, the blurred reflection  model provided statistically
  the best-fit to all the four observations. Irrespective of the
  models used, the soft X-ray excess emission requires contribution
  from a thermal component similar to that expected from an accretion
  disk. The partial covering absorption model results in a nonphysical
  high temperature ($kT_{in} \sim 100\ev$) for an accretion disk and
  is also statistically the worst fit among the three
  models. 1H~0419--577 showed remarkable X-ray spectral
  variability. The soft X-ray excess and the power-law both became
  weaker in January 2010 as well as in May 2010. A moderately broad
  iron line, detected in July 2007, is absent in the January 2010
  observation. Correlated variability of the soft X-ray excess and the
  iron $K\alpha$ line strongly suggest reflection origin for both the
  components. However, such spectral variability cannot be explained
  by the light bending model alone and requires changes in the
  accretion disk/corona geometry possibly arising from changes in the
  accretion rate.
\end{abstract}

\section{Introduction}

Active Galactic Nuclei (AGNs) exhibit complex X-ray spectra. Seyfert 1
galaxies generally show three primary components -- power-law
continuum, soft X-ray excess below $2\kev$ and reflection
    including an Fe~K$\alpha$ line. The powerlaw component is
thought to arise due to
% Our current understanding is that the power-law component is
% dominated by inverse Compton process
Comptonisation of soft photon from an accretion disk
% which takes place
in a hot corona, either above and below the accretion disk in a
sandwich configuration or around the central
    super-massive black hole (SMBH) \citep{sunyaev80, haardt1991,
  reynolds03}. The origin of the soft X-ray excess emission is not
clearly understood. It is likely the blurred reflection
    from a partially ionised accretion disk
    \citep{1993MNRAS.261...74R, fab2002a}.  The soft excess could
also arise due to the Comptonization of optical/UV radiation from the
accretion disk in a low temperature, optically thick medium
\citep{magdziarz1998, done2011}.  The Fe~K$\alpha$ line
  near $6\kev$ and the Compton hump in the $\sim10-40\kev$ band are
  together known as the X-ray reflection. The Fe~K$\alpha$ line arises
  due to the photoelectric absorption followed by fluorescent line
  emission. The Compton hump is the result of two competing processes,
  the photoelectric absorption and the Compton scattering of the
  illuminating powerlaw continuum \citep{1988MNRAS.233..475G,
    1988ApJ...335...57L}. The reflection features arising from the
inner disk are strongly modified due to the relativisitc effects near
the SMBH,  thus giving rise to the broad relativistic iron
  line \citep[see e.g.,][]{2000PASP..112.1145F,reynolds03}. In most
cases, Seyfert 1 galaxies exhibit the narrow iron K$\alpha$ line and
the Compton hump from distant optically thick matter such as the
putative cold torus, and sometimes show a broad line
\citep{reeves2007, murphy2009}.
          
The primary X-ray emission of Seyfert 1 galaxies may be affected by
absorption in the neutral and partially ionised material along the
line of sight. Approximately $50\%$ Seyfert galaxies show absorption
features e.g., due to O\Rmnum{7} and O\Rmnum{8}, in their spectra near
$1\kev$ \citep[e.g.,][]{blustin2005, piconcelli2005}. These absorption
lines and edges are due to the presence of the warm ionised matter
which is named as ``warm or ionised absorber''
\citep{helpern84,1991ApJ...381...85T,1993ApJ...411..594N,Kaastra2000}. In
some cases, the absorption lines are found in the Fe~K
band
\citep{2003MNRAS.345..705P,2005A&A...442..461D,risaliti2005,braito2007,2009ApJ...701..493R,2009A&A...504..401C,2010A&A...521A..57T}.
Sometimes emission lines from the warm absorbing clouds are also
observed \citep[e.g.,][]{laha2011}. The absorbing clouds may be
neutral or partially ionised and may obscure the central source
partially \citep[see e.g.,][]{turner2009}. The multiple warm and
partial covering absorbers may modify the primary continuum in a
complex way \citep{pounds2004a, pounds2004b,turner2009,
  maiolino2010}. These absorbers can mimic reflection features in the
X-ray spectrum \citep{tmiller2009, risaliti2009}. Thus the
reflection and absorption play a crucial role in
shaping the X-ray spectrum. The broad iron K$\alpha$ line and the
reflection are the most important signatures currently available to
probe the central engine. Hence it is important to investigate the
presence of the complex absorption and remove its effects on
the reflection features in order to probe the central
engine. Sometimes the complex absorption and reflection
    models both describe the data equally well and it is difficult
to rule out one of these models \citep{miller2008, tmiller2009}. The
variability of different spectral components and the relationship
between them i.e., the broadband X-ray spectral variability can
provide additional constraints that may help unravel
the real physical model. Here we study the broadband spectral
variability of Seyfert 1 galaxy 1H~0419--577 -- an AGN in which the
complex absorption and/or the reflection may be shaping
    the broadband continuum.

1H~0419--577 is a radio-quiet AGN at a redshift $z = 0.104$ and is
optically classified as a broad-line Seyfert 1 galaxy
\citep{briss1987, grupe96, guainazzi1998, turner1999}.  It was
observed by both the {\it Extreme Ultraviolet Explorer (EUVE)}
\citep{marshall1995} and {\it ROSAT Wide Field Camera}
\citep{pye1995}. It has been detected as one of the brightest AGN in
the extreme ultra-violet band. 1H~0419--577 has also shown interesting
spectral variability -- strong steepening ($\Gamma \sim 2.5$) to a
more flat power-law continuum ($\Gamma \sim 1.6$). This variable
spectral form has suggested a strong transition due to a decrease in
accretion rate from quasi- to sub-Eddington rates
\citep{guainazzi1998, turner1999, pounds2004a}. The first \suzaku{}
observation of 1H~0419--577 has been studied by \citet{turner2009} and
\citet{walton2010} in detail. \citet{turner2009} described the
broadband X-ray spectrum of this source by a power-law modified by
multiple partial covering absorption (PCA) components -- ($i$) absorption
column N$_{H}\sim 1.8\times10^{24}{~\rm cm^{-2}}$, covering fraction
C$_{f}\sim 66\%$, and ($ii$) N$_{H}\sim 5.4\times10^{22}{~\rm
  cm^{-2}}$, C$_{f}\sim 16\%$. They could not explain the hard excess
by reflection and claimed that the hard excess is caused by the
Compton thick absorber ($ N_{H}\sim1/1.2\sigma_{T}\ge
1.25\times10^{24}{\rm ~cm^{-2}}$).  Whereas, \citet{walton2010} showed
that the broadband X-ray spectrum obtained from the same \suzaku{}
observation is well described by a complex blurred
    reflection model with a broken emissivity law (emissivity
indices: $q_{in}>8.7$ below a break radius $ r_{br}\sim2.4r_{g}$ and
$q_{out}\sim 5$ above the break radius) and inferred a large
inclination angle ($\sim 55^{\circ}$). Here we investigate these
models using a new \suzaku{} observation and two new \xmm{}
observations in addition to the first \suzaku{} observation already
studied by \citet{turner2009} and \citet{walton2010}.  In section 2,
we describe the \suzaku{} and \xmm{} observations and data
reduction. We present spectral modelling in section 3. Finally, we
discuss our results in section 4, followed by conclusions in section
5.  We used the cosmological parameters $H_{0} = 71~{\rm
  km~s^{-1}~Mpc^{-1}}$, $\Omega_m = 0.27$ and $\Omega_{\Lambda} =
0.73$ to calculate the distance.

\begin{table*}
  \centering
  \caption{\suzaku{} and \xmm{} X-ray observations of 1H~0419--577 } \label{obs_log}
  \begin{tabular}{lccccc}
    \hline
    \hline 
    Observatory & Obs. ID  &  Date   & Instrument   & Net exposure   & Rate~~$^{a}$     \tabularnewline
    &          &       &              &(\ks)           & ${\rm counts~s^{-1}}$ \tabularnewline

    \hline
    \suzaku{}      & 704064010     &Jan. 16--18, 2010 & XIS0        & 123  & $0.91\pm{0.003}$     \tabularnewline
    &                  &          & HXD/PIN    & 105       & $(3.8\pm0.2)\times10^{-2}$   \tabularnewline
    ''          & 702041010       &July 25--28, 2007 & XIS0 & 205          &$ 1.4\pm{0.003}$ \tabularnewline
    &                  &          & HXD/PIN     &143          &$(4.8\pm 0.2)\times 10^{-2}$ \tabularnewline
    \xmm{} & 0604720301 & May 30--31, 2010 & EPIC-pn  &107 & $11.86\pm0.02 $ \tabularnewline
    '' & 0604720401 & May 28--29, 2010 & EPIC-pn &61 & $10.88\pm0.02 $    \tabularnewline 

    \hline
  \end{tabular}\\
  Note-- Count rates for XIS0 and HXD/PIN are quoted in the 
  $0.6-10\kev$ band and $15-50\kev$ band, respectively.
\end{table*}

\section{Observation and data reduction}
\subsection{ \suzaku{} observations}
1H~0419--577 has been observed twice with the \suzaku{} X-ray
observatory. The first observation was performed in July 2007 for an
exposure time of $205\ks$ and the second observation was performed in
January 2010 for an exposure time of $\sim123\ks$ (see
Table~\ref{obs_log}). The first observation has been analysed by
\citet{turner2009} and \citet{walton2010}. Here, we present a detailed
analysis of the second new \suzaku{} and recent \xmm{} observations
and compare with the first \suzaku{} observation.
             
We reprocessed XIS and PIN data using the software {\tt HEASOFT v6.12}
and the recent calibration data following the \suzaku{} ABC guide
(v3.2)
\footnote{http://heasarc.gsfc.nasa.gov/docs/suzaku/analysis/abc/}. We
used {\tt aepipeline} to reprocess and filter the unfiltered event
lists and created the cleaned event files. We used {\tt xselect} to
extract spectral products for each of the XIS camera. We extracted
source spectra from a circular region with a radius of $260{\arcsec}$
centred on the source position. We also extracted the background
spectra using circular regions with radii in the range
$83-123{\arcsec}$ avoiding the source and the chip corner where the
calibration sources are registered. We generated the ancillary
response and the redistribution matrix files for each XIS spectral
dataset using {\tt xissimarfgen} and {\tt xisrmfgen}.
  
The HXD is a collimating rather than an imaging instrument and the
estimation of background requires the non X-ray instrumental
background (NXB) and cosmic X-ray background (CXB). We used {\tt
  hxdpinxbpi} to create source and combined NXB+CXB background
spectra. This script requires cleaned event file, the pseudo event
file, and the NXB file. We obtained the tuned background file for this
observation provided by the \suzaku{}
team\footnote{ftp://legacy.gsfc.nasa.gov/suzaku/data/background/pinnxbver2.0tuned/}. This
script also applies dead time correction to the source spectrum using
{\tt hxddtcorr}.
      
We grouped the XIS and PIN spectral data in order to use the
$\chi^{2}$ statistics in our spectral fittings. The XIS spectra were
grouped to result in $\sim 250$ bins so that there are about three
energy bins per resolution element of size $\sim 100\ev$. The PIN
spectra were grouped to result in $\sim 50$ energy bins. Our grouping
scheme ensured that there are more than $20$ counts per bin in the
grouped spectra.

\begin{figure}
  \centering
  \includegraphics[height=8.4cm,angle=-90]{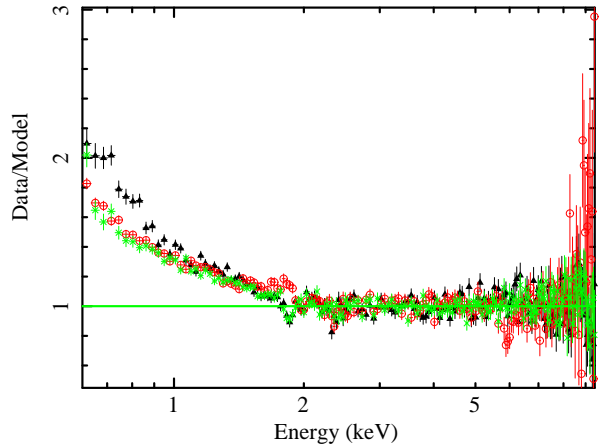}
  \caption{The ratio of the observed spectral data -- XIS0 (solid
    triangles), XIS1 (open circles) and XIS3 (crosses), and the
    $2-10\kev$ absorbed power-law model extrapolated down to $0.6\kev$
    for the January 2010 observation. }
  \label{jan10_spec_comp}
\end{figure}

\begin{figure}
  \centering
  \includegraphics[height=12.0cm]{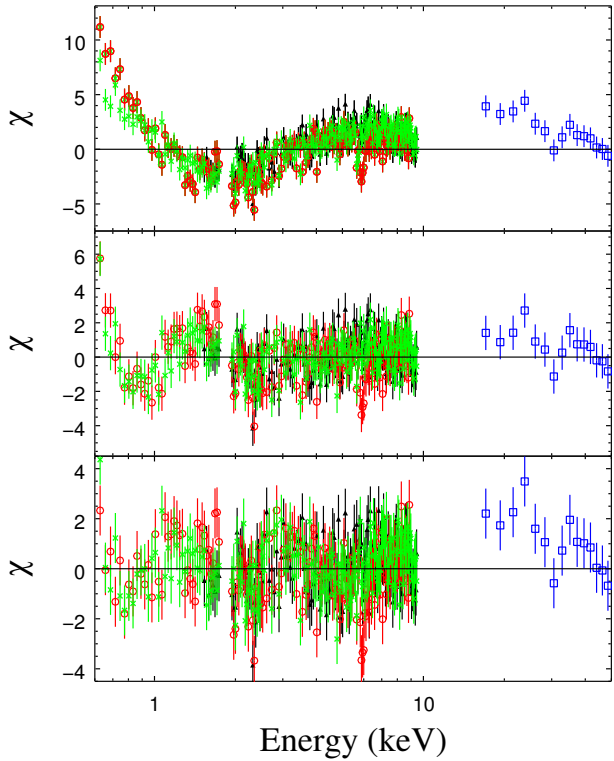}
   \caption{The deviations of the observed data XIS0 (solid triangle),
    XIS1 (open circle), XIS3 (cross) and PIN (open square), from the
    best-fitting models {\tt wabs(powerlaw+zgauss)} ({\it upper
      panel}), {\tt wabs(powerlaw+zbbody+zgauss)} ({\it middle panel})
    and {\tt wabs$\times$ zpcfabs(powerlaw+zbbody+zgauss)} ({\it lower
      panel}) in the $0.6-50\kev$ band. }
  \label{jan10_spec_res}
\end{figure}

\subsection{ \xmm{} May 2010 observation} \xmm{} has observed
1H~0419--577 nine times. The previous seven observations have been
studied by various authors \citep{tombesi2012, tombesi2010,
  walton2010, turner2009, fabian2005, pounds2004a, pounds2004b,
  page2002}. The latest two observation were performed in May 2010
(observation ID: 0604720301 and 0604720401) (see Table~\ref{obs_log})
with exposure times of $106.7\ks$ and $61\ks$. The EPIC-pn camera
\citep{turner2001} was operated in the small window mode using the
thin filter. The two MOS cameras \citep{struder2001} were operated in
the small window mode using the medium filter. Here we analysed the
data obtained from these two observations in 2010 (see
Table~\ref{obs_log}).

We performed standard processing using the \xmm{} Science Analysis
System ({\tt SAS v12.0.1}) and most recent calibration files. We
reprocessed only the EPIC-pn data with {\tt epproc} and obtained event
files. We examined the presence of flaring particle background by
extracting lightcurves above $10\kev$ and found that the first
observation was affected by the flaring particle background. We
excluded the interval of high particle background from the rest of the
analysis. This resulted in the net exposure of $43.2\ks$ for
observation ID 0604720301.  Similarly we found the net exposure of
$42.3\ks$ for the second observation (observation ID 0604720401). We
used single and double events (PATTERN $\le4$) for the EPIC-pn and
omitted events in the bad pixels and next to CCD edges.  We extracted
the source spectrum from a circular region with radius $50{\arcsec}$
for each observation. We also obtained background spectra from the
source-free circular regions with radii in the
$20{\arcsec}-25{\arcsec}$ range.  We generated response matrix and
ancillary response files at the source position using the tools {\tt
  rmfgen} and {\tt arfgen}, respectively. We grouped the EPIC-pn data
appropriately using the SAS task {\tt specgroup} with an oversampling
of $10$ and minimum counts of $20$ per bin.

\begin{table}
  % { \small
  \centering 
  \caption{Best fit parameters of powerlaw plus Gaussian line model fitted to the $2.5-10\kev$ \suzaku{} XIS and \xmm{} EPIC-pn spectra. 
    \label{par_pl1g} }
  \begin{tabular}{lcccc}

    \hline
    \hline 
    & \multicolumn{2}{c}{\suzaku{}} & \multicolumn{2}{c}{\xmm{}} \\
    Parameter                 & January 2010            &  July 2007         & May 2010         & May 2010 \\
    &  (704064010)            &(702041010)           & (0604720301)        & (0604720401)           \\

    \hline
    $ \Gamma$                      & $1.77\pm{0.02}$            & $1.78\pm{0.01}$     &$1.67\pm{0.03}$  &$1.64\pm{0.03}$    \tabularnewline
    \rm $ E_{FeK\alpha}$(keV)          & $6.35_{-0.04}^{+0.05}$     & $ 6.35\pm{0.11}$    &$6.30\pm{0.15}$  &$6.35_{-0.32}^{+0.34}$    \tabularnewline
    $ \sigma $ (\kev)               & $<0.12$                   & $0.37_{-0.14}^{+0.17}$     &$0.2_{-0.1}^{+0.3}$  &$0.4\pm{0.3}$    \tabularnewline
    $ f_{FeK\alpha}$~~$^{a}$       &$0.5\pm{0.2}$      & $1.3_{-0.4}^{+0.5}$     &$1.0_{-0.5}^{+0.9}$  &$0.9_{-0.6}^{+0.7}$    \tabularnewline
    $ f_{PL}$~~$^{b}$   &$1.47_{-0.02}^{+0.01} $     & $1.76_{-0.02}^{+0.01}$     &$1.29_{-0.01}^{+0.03}$  &$1.26\pm{0.01}$    \tabularnewline
    $ \chi^{2}/dof$                &$367.2/343$                &$386.2/353$                &$310.7/295$    &$189.5/176$   \tabularnewline
    $\Delta\chi^2$ ~~$^{c}$        &   $-18.5$           & $-46.9$         & $-15.4$     & $-8.1$  \tabularnewline
   
    \hline
  \end{tabular} \\
  Note -- (a) Iron K$\alpha$ line flux in units of $10^{-5}$ photons $\rm cm^{-2}~s^{-1}$; (b) $2-10\kev$ power-law flux in units of $10^{-11}$ erg $\rm cm^{-2}~s^{-1}$; (c) Represents reduction in $\chi^{2}$ with extra three parameters for iron K$\alpha$ line.
\end{table}

Based on our spectral analysis below, we found that the $2-10\kev$
flux and luminosity of 1H~0419--577 were $F_{2-10 \kev}\sim 1.4
\times10^{-11}$ erg $\rm cm^{-2}~s^{-1}$ and $ L_{2-10
  \kev}\sim10^{45}$~$\rm erg~s^{-1}$, respectively, during the January
2010 \suzaku{} and May 2010 \xmm{} observations. This flux represents
the low state of the source as compared to July 2007 observation
\citep{page2002, pounds2004a, turner2009}.

\begin{figure}
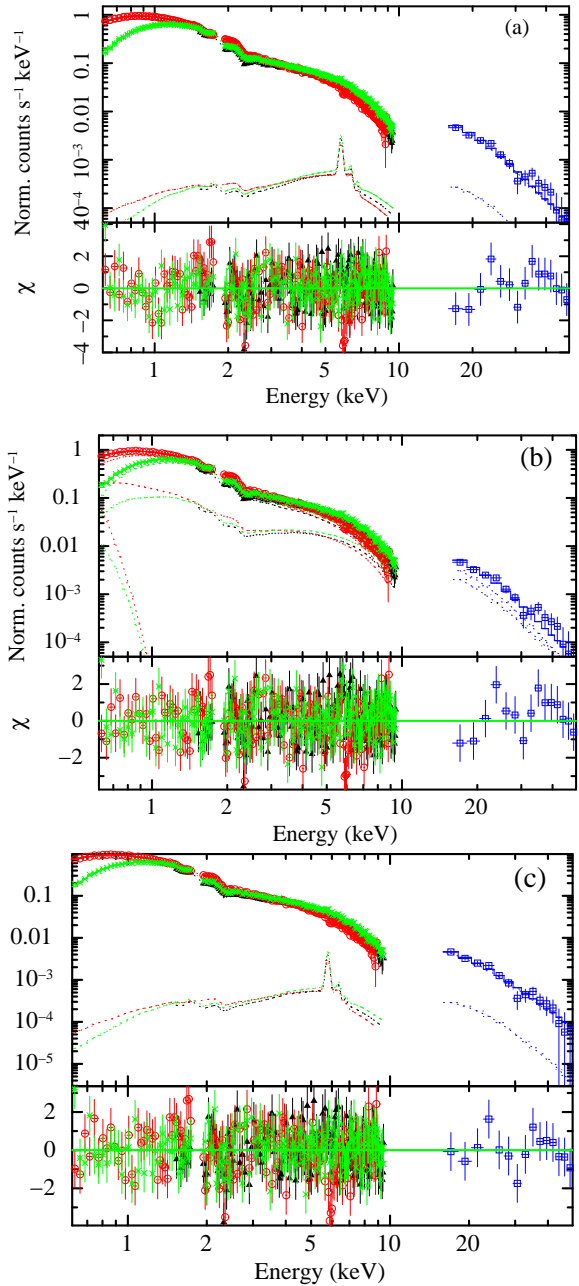

  \centering
  \includegraphics[width=5.8cm,angle=-90.00]{fig3a.ps}
  \includegraphics[width=5.8cm,angle=-90.00]{fig3b.ps}
  \includegraphics[width=5.8cm,angle=-90.00]{fig3c.ps}
  \caption{Results of spectral analysis of the January 2010 \suzaku{}
    observation. The observed XIS+PIN spectral data, the best-fit
    model and the deviations.  The best-fit models are (a) the complex
     ionised PCA model ({\tt
          wabs$\times$zxipcf$\times$zxipcf(powerlaw+pexmon)}),
    (b) the blurred reflection model ({\tt
          wabs(nthcomp+kdblur{*}reflionx+diskbb)}), and (c) the
    intrinsic disk Comptonization model ({\tt wabs(optxagnf+pexmon)}).
    The symbols used for different data are the same as in
    Figure~\ref{jan10_spec_res}.}
  \label{jan10_fit_fig}

\end{figure}

\begin{table*}
  \small
  \centering 
  \caption{Best fit parameter for \suzaku{} and \xmm{} observations, and Fixed parameters are indicated by an asterisk.} \label{pcaref_fit_par}

  \begin{tabular}{llccccc}
    \hline
    \hline 
    \multicolumn{6}{c}{Model 1: Ionised PCA model}\\ 
    \hline
    &     & \multicolumn{2}{c}{\suzaku{}} & \multicolumn{2}{c}{\xmm{}} \\
               
    % \hline

    Model&Parameter&704064010~(January 2010)&702041010~(July 2007)&0604720301~(May 2010)&0604720401~(May 2010)\\
    Component &     &  &  &  & \\
    \hline
    Gal. abs.&$N_{H}$ ($10^{20}\rm cm^{-2}$) & 1.83~(*)& 1.83~(*)&1.83~(*) &1.83~(*)\tabularnewline
  
    Powerlaw&$\Gamma$&$2.27_{-0.03}^{+0.02}$&$2.33\pm0.01$&$2.52\pm0.03$&$2.44\pm0.03$\tabularnewline
    &n$_{pl}~(10^{-2})$~$~^{a}$&$2.0\pm0.2$ &$3.1\pm0.2$&$2.9\pm0.3$&$2.6\pm0.2$\tabularnewline    
    Zxipcf~(1)& $N_{H}$ ($10^{23}{\rm cm^{-2}}$) &$5.1_{-0.8}^{+0.7}$&$4.6\pm0.4$&$4.2_{-0.8}^{+0.5}$ &$4.5_{-0.8}^{+0.5}$\tabularnewline
    & $C_{f}$ ($\%$) &$50.0_{-5.0}^{+2.4}$&$60.0_{-2.6}^{+2.2}$ &$66.1_{-3.4}^{+2.4}$ &$67.2_{-2.3}^{+1.7}$\tabularnewline
    & $\xi$ ($\rm erg~cm~s^{-1}$)&$0.28_{-0.17}^{+0.37}$&$0.06_{-0.04}^{+0.04}$ &$1.4_{-0.8}^{+1.1}$&$1.4_{-0.8}^{+1.0}$\tabularnewline   
    Zxipcf~(2)$~~^{b}$& $N_{H}$ ($10^{23}\rm cm^{-2}$) &$1.4\pm0.1$&$1.3_{-0.12}^{+0.06}$&$0.53_{-0.05}^{+0.06}$&$0.58_{-0.05}^{+0.06}$\tabularnewline
    &$C_{f}$ ($\%$) &$46.6_{-2.9}^{+1.3}$&$46.9_{-4.0}^{+3.6}$&$52.7_{-2.6}^{+2.3}$&$50.3_{-2.7}^{+2.5}$\tabularnewline
    & $\xi$ ($\rm erg~cm~s^{-1}$)&$88.2_{-5.8}^{+7.9}$&$77.8.3_{-26.1}^{+4.6}$ &$20.0_{-3.7}^{+6.5}$&$23.4_{-4.8}^{+7.5}$\tabularnewline
 %   $FeK\alpha$ &$E_{FeK\alpha}$&$6.35\pm0.05$&$6.36_{-0.07}^{+0.06}$&--&-- \tabularnewline
 %   &$\sigma~{(\kev)}$&$0.01~(*)$&$0.01~(*)$&--&--  \tabularnewline
 %   &f$_{FeK\alpha}~(10^{-6})$~$~^{a}$&$8.3_{-3.9}^{+4.1}$&$8.0_{-3.8}^{+4.9}$&--&-- \tabularnewline
 %   &EW~$(\ev)$&$18.6$&$11.9$&--&-- \tabularnewline
 %   &$\Delta \chi^{2}$&$-12.1$&$-14.2$&--&-- \tabularnewline

Pexmon & $R$ & $-0.13_{-0.12}^{+0.07}$ & $-0.12\pm0.06$ & $-0.10_{-0.19}^{+0.10}$ & $-0.002_{-0.16}^{+0.002}$\\

    &$\chi^{2}/\nu$&$613.3/500$ & $786.9/541$&$560.5/423$ & $585.0/419$\tabularnewline
    $F_{X}~~^{c}$&$f_{0.6-2keV}$&$0.98$&$1.22$&$0.90$&$0.82$  \tabularnewline
    &$f_{2-10keV}$&$1.47$&$1.79$&$1.31$&$1.28$\tabularnewline
    &$f_{10-50keV}$&$2.4$&$3.3$&--&--\tabularnewline
    $L_{X}~~^{d}$&$L_{0.6-2keV}$&$2.8$&$3.5$&$2.6$&$2.3$ \tabularnewline
    &$L_{2-10keV}$ &$4.0$&$4.8$&$3.5$&$3.4$\tabularnewline
    % &$L_{10-50keV}$&$6.6$&$8.9$&--&--\tabularnewline
    \hline
    \multicolumn{6}{c}{Model 2: Blurred reflection model}\\ 
    \hline
    Nthcomp$~^{e}$ & $\Gamma $&$2.06_{-0.02}^{+0.03}$&$2.07\pm{0.01}$ &$2.12\pm{0.02}$&$2.08\pm{0.01}$  \tabularnewline
    &$kT_{in}$~(\ev)& $36.0_{-19.6}^{+17.9}$&30~(*)&$25.6_{-6.7}^{+7.0}$&$23.18\pm{0.01}$ \tabularnewline
    &n$_{nth}~(10^{-3})~~^a$~&$4.3_{-0.2}^{+0.5}$&$5.48_{-0.15}^{+0.02}$&$3.8\pm{0.2}$&$3.5\pm{0.1}$ \tabularnewline
    Diskbb&$kT_{in}$ (\ev) &$36.0_{-19.6}^{+17.9}$&--&$25.6_{-6.7}^{+7.0}$&$23.18\pm{0.01}$ \tabularnewline
    &n$_{diskbb}~(10^{8})$~$~~^{a}$&$>0.001$&--&$2.14_{-2.03}^{+14.49}$&$6.21_{-5.95}^{+87.65}$\tabularnewline      
    Reflionx(1) & $ A_{Fe}$&$0.5_{-0.1}^{+0.2}$&$0.6\pm{0.1}$&$0.4\pm{0.1}$&$0.3_{-0.2}^{+0.1}$\tabularnewline
    &$\xi$ ($\rm erg~cm~s^{-1}$)&$37.3_{-23.1}^{+30.6}$&$197.9_{-79.6}^{+7.5}$&$25.2_{-4.1}^{+7.9}$&$21.0_{-5.8}^{+2.0}$\tabularnewline
    &$\Gamma$ &$2.06_{-0.02}^{+0.03}$&$2.07\pm{0.01}$ &$2.12\pm{0.02}$&$2.08\pm{0.01}$\tabularnewline
    &n$_{ref}~(10^{-6})$~$~^{a}$ & $8.2_{-1.7}^{+4.9}$&$0.6_{-0.1}^{+0.4}$&$14.0_{-4.0}^{+3.0}$&$15.0_{-3.0}^{+2.0}$\tabularnewline
    Reflionx(2) & $ A_{F_{e}}$ &--& $0.6\pm{0.1}$&--&-\tabularnewline
    &$\xi$~($\rm erg~cm~s^{-1}$)&--& $<11$ &--&--\tabularnewline
    &$\Gamma$ &--&$2.07\pm{0.01}$ &--&--\tabularnewline
    &n$_{ref}~(10^{-5})$~$~^{a}$ &-- &$2.6_{-0.4}^{+0.2}$&--&--\tabularnewline

    Kdblur(1)& $ q $ & $8.5\pm{0.9}$& $>8.6$&$>6.04$&$8.4_{-2.2}^{+0.7}$\tabularnewline
    & $R_{in}$ ($r_{g}$)& $1.47_{-0.08}^{+0.22}$&$1.56\pm{0.05}$&$1.4\pm{0.1}$&$1.4_{-0.1}^{+0.2}$\tabularnewline
    & $R_{out}$ ($r_{g}$)& 400~(*) &$2.2_{-0.2}^{+0.1}$ &$400~(*)$&$400~(*)$\tabularnewline
    & $i$ ($\rm degree$)&$50.4_{-8.3}^{+5.9}$ &$51.3_{-1.3}^{+2.9}$&$57.8_{-7.6}^{+4.4}$&$60.0_{-7.8}^{+1.9}$\tabularnewline 
    Kdblur(2) & $ q_{out}$&--& $5.0_{-0.4}^{+0.9}$&--&--\tabularnewline
    &$R_{br}$ ($r_{g}$)&--&$2.2_{-0.2}^{+0.1}$&--&--\tabularnewline
    &$R_{out}$ ($r_{g}$)&--& 400~(*)&--&--\tabularnewline
    &$i$ ($\rm degree$)&--& $51.3_{-1.3}^{+2.9}$&--&--\tabularnewline   
    &$ \chi^{2}/\nu$&$580.5/499$&$735.2/538$&$460.8/422$&$447.7/418$\tabularnewline
    $ F_{X}~~^{c}$&$f_{0.6-2keV}$ &$0.98$&$1.2$&$0.9$&$0.8$  \tabularnewline
    & $f_{2-10keV}$ &$1.45$&$1.77$&$1.30$&$1.27$\tabularnewline
    & $f_{10-50keV}$ &$2.5$&$3.0$&--&--\tabularnewline
    $ L_{X}~~^{d}$&$L_{0.6-2keV}$&$2.7$&$3.4$&$2.5$&$2.3$ \tabularnewline
    &$L_{2-10keV}$&$3.9$&$4.7$&$3.4$&$3.3$\tabularnewline
    % & $L_{10-50keV}$ &$6.7$&$8.1$&--&--\tabularnewline

    \hline
  \end{tabular}
  {~~~~~~~~~~~~~~~~~~~~~~~~~~~~~~Notes-- (a) n$_{pl}$, n$_{nth}$, n$_{diskbb}$ and n$_{ref}$ represent normalization to respective model component; where n$_{pl}$ and n$_{pex}$ have same units as ${\rm~photons~keV^{-1}~cm^{-2}~s^{-1}}$ at 1 keV;  (b) ionised PCA (1 or 2); (c) Flux in units $10^{-11}{\rm~ergs~cm^{-2}~s^{-1}}$;
    (d) Luminosity in units $10^{44}{\rm~ergs~s^{-1}}$; (e) Electron temperature is fixed to 100 $\kev$.}
\end{table*}

\section{Spectral modelling}
We used {\tt XSPEC v12.7.1} to analyse the \suzaku{} and \xmm{} X-ray
spectra of 1H~0419--577. We used the $\chi^2$ minimisation technique
to find the best-fit models. Below we quote the errors on the best-fit
parameters at the $90\%$ confidence level unless otherwise specified.
\subsection{January 2010 \suzaku{} data}
First we compared XIS spectra to find if there are any
instrumental cross-calibration issues. We fit an absorbed power-law
model ({\tt constant$\times$ wabs$\times$ powerlaw}) jointly to three
XIS (XIS0, XIS1 and XIS3) spectra in $2-10\kev$ band.  We fixed the
constant to 1 for XIS0 and varied for the XIS1 and XIS3 datasets in
order to account for any variations in the relative normalisation of
the three detectors. We fixed the column density at the Galactic value
of $N_{H}= 1.83\times10^{20}$~$\rm cm^{-2}$ \citep{dickey1990}.  The
ratio of the observed XIS data and the best-fit absorbed power-law
model extrapolated to lower energies is shown in
Figure~\ref{jan10_spec_comp}.  It is clear that the XIS0 data does not
agree with the XIS1 and XIS3 data below $1.5\kev$. We therefore
excluded the XIS0 data below $1.5 \kev$. XIS1 is affected by high
background near $10\kev$ and therefore we ignored XIS1 data above
$8.9\kev$. Moreover, all XIS datasets show the calibration
uncertainties near Si-K edge which led to the exclusion of
$1.78-1.9\kev$ spectral range. We also excluded any bad channels from
the spectral modelling.

We began with the spectral modelling of the $2.5-10\kev$ XIS data to
find the hard power-law component. An absorbed power-law model
resulted in a minimum $\chi^{2}=385.7$ for 346 degree of freedom
(dof). We noticed a narrow emission line near $6\kev$. Addition of a
redshifted Gaussian line improved the fit by $\Delta \chi^{2} = -18.5$
for three additional parameters. The width of the line ($\sigma$) was
consistent with zero. The $90\%$ upper limit on the Gaussian $\sigma$
is $120\ev$. We found an equivalent width of $24.7\ev$ for this narrow
K$\alpha$ line. The best-fit powerlaw photon index is $\Gamma \sim
1.8$ in the $2.5-10\kev$ band. The best-fit parameters are listed in
Table~\ref{par_pl1g}. Next, we fitted the absorbed power-law plus
narrow iron line model to the $0.6-10\kev$ XIS data and the
$15-50\kev$ HXD/PIN data jointly.  We fixed the relative normalisation
of the PIN data at $1.18$ appropriate for this observation at the HXD
nominal position\footnote{\url{
    http://heasarc.gsfc.nasa.gov/docs/suzaku/analysis/watchout.html}}.
The fit resulted in $\chi^{2}/dof= 2202.1/505$. We show the residuals
in Figure~\ref{jan10_spec_res} (a). It is clear that the broadband
continuum of 1H~0419--577 consists of a power-law, soft X-ray excess
below $\sim 1\kev$ and possible hard X-ray excess above $\sim 10\kev$.

\begin{figure*}
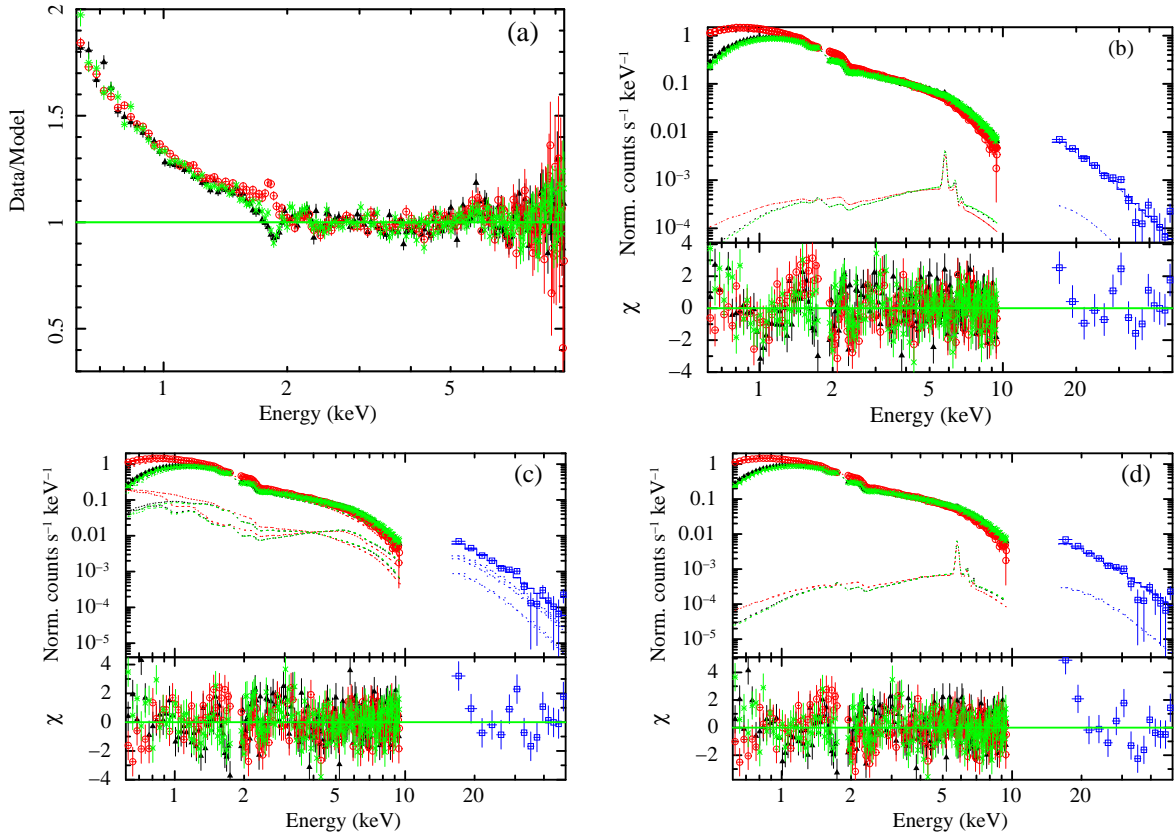

  \centering
  \includegraphics[height=8cm,angle=-90.00]{fig4a.ps}
  \includegraphics[height=8cm,angle=-90.00]{fig4b.ps}
  \includegraphics[height=8cm,angle=-90.00]{fig4c.ps}
  \includegraphics[height=8cm,angle=-90.00]{fig4d.ps}
  \caption{Results of the spectral analysis of July 2007 \suzaku{}
    observation.  (a) A comparison of XIS0, XIS1 and XIS3 spectral
    data in terms of the ratio of the observed data and the best-fit
    $2-10\kev$ absorbed powerlaw ($\Gamma\sim 1.8$) model extended to
    low energies.  XIS+PIN spectral data, the best-fitting models and
    the deviations. The best-fitting models are (b) the complex
    ionised PCA, (c) the blurred reflection, and (d)
    the intrinsic disk Comptonization model.}
  \label{suzaku07_spec_fit}

\end{figure*}

The soft excess is a common feature of type 1 AGNs
\citep[e.g.,][]{boller1996,leighly1999,vaughan1999,gier2004,crum2006})
and its origin is still unclear. Several models such as single
blackbody, multiple blackbodies, multicolor disk blackbodies, blurred
disk reflection from partially ionised material, smeared absorption,
and thermal Comptonization in optically thick medium can provide
statistically good fit to the observed soft excess
\citep{magdziarz1998, fabian2002, gier2004,crum2006,dewangan2007}. We
used a simple blackbody model to describe the soft excess emission of
1H~0419--577. The addition of the blackbody component to earlier
absorbed powerlaw plus narrow Gaussian line model improved the fit
from $ \chi^{2}/dof = 2202.1/505$ to $ \chi^{2}/dof = 774.4/503$ for
two additional parameters. The best-fit blackbody temperature is $\sim
0.17\kev$. We show the fit-residuals in Figure~\ref{jan10_spec_res}
({\it middle panel}).

\subsubsection{Partial covering absorption (PCA) model}
The blackbody plus powerlaw model poorly describes the observed
spectrum. There are features in the residuals below $2\kev$ and slight
spectral curvature in the $2-5\kev$ (see Fig.~\ref{jan10_spec_res},
middle panel). These features may arise due to the presence of PCA. We
used a neutral PCA model {\tt zpcfabs}. This model describes
absorption by neutral cloud that may cover the source partially. The
model {{\tt wabs$\times$ zpcfabs(zbbody + powerlaw + zgauss)}}
improved the fit to $ \chi^{2}/dof = 671.4/501$ ($ \Delta\chi^{2} =
-103.0$ for two additional parameters). However, the fit is still not
satisfactory as the fit-residuals show ``hard excess" above $10\kev$
as shown in Figure~\ref{jan10_spec_res} ({\it lower panel}). This hard
excess could arise due to the presence of another partial covering
absorber as suggested by \citet{turner2009}. Using a second PCA i.e.,
the model {{\tt wabs$\times$ zpcfabs(1)$\times$zpcfabs(2)(zbbody +
    powerlaw + zgauss)}} (the neutral PCA model), improved the fit
further ($\chi^{2}/dof = 618.3/499$).  Using a third neutral PCA did
not improve the fit ($\Delta \chi{2}=0.05$ for two additional
parameters). We therefore find a good fit with two PCAs. However a
{\tt blackbody} model is not a physical model for the soft X-ray
excess emission. We therefore tested if ionised PCA can cause the
spectral curvature.  Turner et al. (2009) used two partially ionised
PCA models to describe the 2007 \suzaku{} observation of
1H0419--577. We constructed a similar model for the 2010 \suzaku{}
data.  We used the ({\tt zxipcf}) model for the ionised absorption.
This model uses a grid of XSTAR photionised absorption models
calculated assuming a gas tubulent velocity of $200{\rm~km~s^{-1}}$, a
powerlaw ($\Gamma=2.2$) illuminating continuum and solar abundances
\citep{reeves2008}.  We used an ionised absorber for the low energy
spectral curvature and another ionised absorber for the hard
excess. Thus, we fitted the ionised PCA model {\tt
  wabs$\times$zxipcf$\times$zxipcf$\times$powerlaw} resulted in
$\chi^{2}/dof = 620.9/501$. A narrow line feature near 6 $\kev$ was
modeled by using a {\tt Gaussian} with $\sigma=0.01\kev$. This further
improved the fit to $\chi^{2}/dof = 608.8/499$. The
best-fit iron line parameters are $E_{FeK\alpha}=6.35\pm0.05\kev$,
  $f_{FeK\alpha} =
  8.3_{-3.9}^{+4.0}\times10^{-6}{\rm~photons~cm^{-2}~s^{-1}}$ and
  equivalent width, $EW=18.6\ev$. The neutral and narrow iron line is
  thought to be associated with the distant, cold
  reflection. Therefore, we used the {\tt pexmon} model to describe
  both the iron line and the reflection hump. We set the pexmon model
  to produce the reflection component only by making the relative
  reflection parameter ($R$) negative, and tying the photon index and
  the normalisation with the corresponding parameters of the primary
  powerlaw. We fixed the high energy cutoff of the illuminating
  powerlaw at $100\kev$ and the abundances of the distant reflector to
  the solar values. We also fixed the inclination angle to
  $50{\rm~degrees}$. Thus, our final ionised PCA model {\tt
    wabs$\times$zxipcf$\times$zxipcf(powerlaw$+$pexmon)} resulted in
  $\chi^2/dof=613.3/500$. The spectral data, the best-fitting complex
ionised PCA model and deviations are shown in
Figure~\ref{jan10_fit_fig}(a).  The best-fit parameter are listed in
Table~\ref{pcaref_fit_par}. The quoted errors are at the $90\%$
confidence level.

\subsubsection{Blurred reflection model}                        
The soft X-ray excess below $2\kev$ and the hard excess above $10\kev$
can also be described as reflection from partially ionised
disk. \citet{walton2010} have shown that the broadband spectrum of
1H~0419--577 obtained from the first \suzaku{} 2007 observation can be
well described by the blurred reflection model. Following
\citet{walton2010}, we fit the 2010 \suzaku{} observation with a
reflection model.  We used the {\tt reflionx} model which describes
the reflection from partially ionised accretion disk \citep{ross2005}.
In this model, the lamp post geometry is assumed in which a compact
corona located above the black hole illuminates the accretion disk
resulting in radius dependent irradiation. Thus, the emissivity
associated with the reflection depends on the radial distance and
parameterised as $\epsilon(r) \propto r^{-q}$. In this geometry, the
irradiation decreases as $r^{-3}$ far away from the black hole but the
light bending affects the illumination strongly in the central regions
and focuses some fraction of the coronal emission.  The emissivity law
can be much steeper in the innermost regions
\citep{2004MNRAS.349.1435M}. Thus, the emissivity law can be
approximated as a broken powerlaw \citep{fabian2012} over the radial
extent of the disk.

\begin{table*}
  \small
  \centering 
  \caption{Best fit parameter for \suzaku{} and \xmm{} observations, and {*} is used for fixed parameters.} \label{pexopt_fit_par}
  \begin{tabular}{llccccc}
    \hline
    \hline 
    \multicolumn{6}{c}{Model 3: Intrinsic disk Comptonization model}\\ 
    \hline
    &     & \multicolumn{2}{c}{\it \suzaku{}} & \multicolumn{2}{c}{\it \xmm{}} \\               
    % \hline
    Model&Parameter&704064010~(January 2010)&702041010~(July 2007)&0604720301~(May 2010)&0604720401~(May 2010) \\
    Component &     &  &  &  & \\
    \hline
    Gal. abs.&$N_{H}~(10^{20}\rm cm^{-2})$&$1.83~(*)$&$1.83~(*)$&$1.83~(*)$&$1.83~(*)$\tabularnewline
    powerlaw &$\Gamma$&$1.75_{-0.03}^{+0.02}$&$1.67_{-0.04}^{+0.06}$&$1.63_{-0.06}^{+0.05}$&$1.66\pm{0.02}$\tabularnewline
    &n$_{pl}~(10^{-2})$~$~^{a}$&$0.38_{-0.02}^{+0.01}$&$0.45\pm{0.01}$&$0.278_{-0.004}^{+0.002}$&$0.29\pm{0.01}$ \tabularnewline

    Pexmon~$~~^{b}$&$\Gamma$&$1.76\pm{0.02}$&$1.73_{-0.03}^{+0.02}$&$1.66_{-0.07}^{+0.05}$&$1.55_{-0.05}^{+0.07}$\tabularnewline
    & R & $-0.13\pm0.05$&$-0.13_{-0.03}^{+0.04}$&$-0.15_{-0.09}^{+0.08}$&$-0.09\pm0.06$\tabularnewline
    &$n_{pex}~(10^{-3})$~$~^{a}$&$3.8~(*)$&$4.5~(*)$&$2.78~(*)$&$2.9~(*)$  \tabularnewline

    optxagnf$~~^{c}$
    &$L/L_{Edd}$ &$0.80_{-0.05}^{+2.97}$&$0.78_{-0.06}^{+0.08}$&$0.58_{-0.05}^{+0.16}$&$0.61_{-0.57}^{+0.24}$\tabularnewline
    &$kT_{e}$ (\kev)&$0.30_{-0.03}^{+0.05}$&$0.56_{-0.08}^{+0.11}$&$0.6\pm0.1$&$1.2_{-0.23}^{+0.51}$\tabularnewline
    &$\tau$ &$16.0_{-3.2}^{+4.9}$&$8.2_{-0.6}^{+1.1}$&$7.7_{-0.7}^{+0.6}$&$5.3_{-0.8}^{+1.0}$\tabularnewline
    &$r_{cor}~(r_{g})$&$2.1_{-0.1}^{+2.3}$&$2.5_{-0.1}^{+0.6}$&$18.5_{-7.0}^{+17.7}$&$23.3_{-10.5}^{+32.5}$\tabularnewline
    &$a$&$>0.923$&$0.988_{-0.003}^{+0.007}$&$0.9~(*)$&$0.9~(*)$\tabularnewline
    &$f_{pl}$ &$0.90_{-0.03}^{+0.02}$&$0.80\pm{0.02}$&$0.42_{-0.21}^{+0.14}$&$0.47_{-0.30}^{+0.20}$\tabularnewline
    &$\Gamma$&$1.76\pm{0.02}$&$1.73_{-0.03}^{+0.02}$&$1.66_{-0.06}^{+0.05}$&$1.55_{-0.05}^{+0.07}$\tabularnewline
 
    &$\chi^{2}/\nu$&$568.7/501$&$750.0/542$&$585.9/425$&$635.0/421$\tabularnewline
    $F_{X}~~^{d}$&$f_{0.6-2keV}$&$0.98$&$1.22$&$0.90$&$0.8$\tabularnewline
    & $f_{2-10keV}$&$1.46$&$1.77$&$1.3$&$1.3$\tabularnewline
    & $f_{10-50keV}$&$2.7$&$3.3$&--&--\tabularnewline
    $L_{X}~~^{e}$&$L_{0.6-2keV}$&$2.7$&$3.3$&$2.5$&$2.3$\tabularnewline
    &$L_{2-10keV}$&$3.9$&$4.7$&$3.4$&$3.4$\tabularnewline
    % & $L_{10-50keV}$&$7.1$&$8.5$&--&--\tabularnewline
    \hline
  \end{tabular}
  {Note-- (a) $n_{pl}$ and $n_{pex}$ represent normalisation to respective model components in units of ${\rm~photons~keV^{-1}~cm^{-2}~s^{-1}}$ at $1\kev$. (b) Iron abundance relative to solar was fixed to unity and the inclination was fixed to $50^\circ$.  (c) Black hole mass, distance and normalisation are fixed to $10^{8}~M_{\odot}$, $474~Mpc$ and unity, respectively. (d) X-ray flux in units of $10^{-11}{\rm~ergs~cm^{-2}~s^{-1}}$.
    (e) X-ray luminosity in units of $10^{44}{\rm~ergs~s^{-1}}$}.
\end{table*}

We began with a single reflection component for broadband
($0.6-50\kev$ band) modified by the Galactic absorption ($
N_{H(G)}=1.83 \times10^{20}$ $\rm cm^{-2}$). In this model, we tied
the shape of the illuminating power-law in the reflection model to
that of the primary power-law continuum. We excluded the narrow {\tt
  Gaussian} line from the fit because the reflection model can take
care of all the emission lines. Initially we did not smooth the
reflected emission due to the relativistic effects. Thus it showed a
number of emission lines below $2\kev$, which are not
    observed.  We noticed here that the high energy curvature is
modeled by the Compton hump. This fit resulted in
$\chi^{2}/dof = 1077.8/504$.  Next we smoothed the reflection
component due to the relativistic effects by using the convolution
model {\tt kdblur} \citep{laor1991, fabian2002}.  We fixed the outer
radius at $\rm R_{out}=400r_{g}$ and inclination angle at $35^{\circ}$
an appropriate value for a type 1 AGN. Remaining parameters were
varied. This improved the fit statistically from $ \chi^{2}/dof
=1077.8/504$ to $ \chi^{2}/dof= 604.4/502$ with two additional
parameters.

We also noticed large residuals at low energies. This suggested that
the soft excess is not well described by the reflection model
alone. The residuals may be suggesting the possible presence of tail
of the thermal accretion disk emission. We therefore added a blackbody
component. We found a significant improvement in the fit with $ \Delta
\chi^{2} = -18.9$ for two additional parameters ($ \chi^{2}/dof =
585.5/500$).

The power-law component is thought to arise due to thermal
Comptonization of disk photons in a hot electron corona. Therefore, we
replaced the power-law by the Comptonization model {\tt nthcomp} and
the {\tt blackbody} by the multicolor disk blackbody model {\tt
  diskbb}. We tied the temperature of the seed photons in the {\tt
  nthcomp} with the temperature of the inner disk in {\tt diskbb}
component. We varied the diskbb normalization. The model, {\tt
  wabs(nthcomp+diskbb+kdblur*reflionx)} (blurred reflection),
resulted in $\chi^{2}/dof = 582.1/500$ similar to the earlier fit as
expected.  We also varied the inclination angle which resulted in $
\chi^{2}/dof = 580.5/499$ and $i=57.8_{-7.6}^{+4.4}$~degrees. This
inclination angle is similar to that obtained by \cite{walton2010} for
the July 2007 \suzaku{} observation.
% based on a complex reflection model fitted to the broadband X-ray
% spectrum of 1H~0419--577 obtained from the July 2007 \suzaku{}
% observation.
We also tested a more complex two component reflection model similar
to that used by \cite{walton2010} for the July 2007 observation. We
used the model {\tt wabs(powerlaw+kdblur*reflionx+ kdblur*reflionx)}
and tied the outer radius of first {\tt kdblur} component to the inner
radius of the second {\tt kdblur} component. We refer this model as
the complex blurred reflection model. Thus, we made the emissivity
index and the ionisation parameter radius dependent. This model did
not result in a statistically significant improvement $\chi^{2}/dof =
576.3/497$. Thus, the complex blurred reflection model with the broken
emissivity law is not required by the 2010 observation.  The observed
data, the best-fitting blurred reflection model, and the deviations
are shown in Figure~\ref{jan10_fit_fig} (b) and the best-fit
parameters are listed in Table~\ref{pcaref_fit_par}.

\begin{figure}
  \centering
  \includegraphics[height=8.4cm,angle=-90.00]{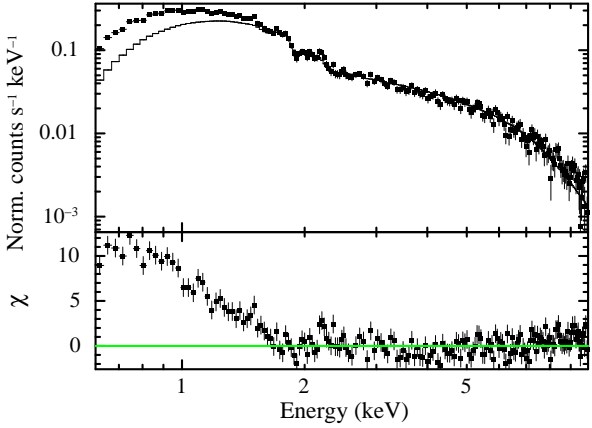}
  \caption{The difference of 2007 and 2010 \suzaku{} XIS0+XIS3
    spectra, best-fitting $2-10\kev$ power-law extrapolated down to
    $0.6\kev$ and the deviations showing the stronger soft excess
    during 2007 observation.}
  \label{diff_spec}
\end{figure}

\subsubsection{Intrinsic disk Comptonization model}

Finally, we modeled the soft excess as the intirinsic disk emission
including Compton upscattering in the disk itself as implemented in
the {\tt optxagnf} model by \cite{done2011}. In this model, the soft
excess can be explained partly by a colour temperature corrected
intrinsic disk emission from the region $r_{out}$ to $r_{cor}$ and
Compton upscattering of this emission in the upper layers of the inner
disk, $r_{cor}$ to $r_{isco}$, consisting of optically thick, low
temperature electron plasma. This model also includes the hard
power-law component arising from the Comptonization of disk emission
in a hot ($\rm kT\simeq 100\kev$), optically thin corona ($ \tau
\simeq 1$) below a radius $r_{cor}$. The hard power-law constitutes a
fraction $f_{pl}$ of the gravitation energy released between $r_{cor}$
and $r_{isco}$, and the remaining fraction $1-f_{pl}$ is emitted as
the optically thick thermal Comptonization contributing to the soft
X-ray excess emission.  In the {\tt optxagnf} model, the soft excess
is the intrinsic disk emission and so its strength is set by the black
hole mass, spin and mass accretion rate through the outer disk
\citep{done2011}.

We started with the absorbed power-law in the full band ($0.6-50\kev$)
as described above. We added XSPEC model {\tt optxagnf} and excluded
the {\it powerlaw} model. In the {\tt optxagnf} model, we used a black
hole mass, $M_{BH}\sim 1\times10^{8}~M_{\odot}$, for 1H~0419--577
estimated by \cite{pounds2004b}.  Thus, we fixed the black hole mass
and the luminosity distance ($d_{L} = 474\mpc$). We also fixed the
normalisation to unity, spin parameter to 0.9 \citep{fabian2005} and
varied $L/L_{Edd}$ to obtain the flux. We fixed $r_{cor}$ to its
best-fit value $15~r_{g}$. The rest of the parameters were varied to
obtain a good fit.  The fit resulted in $ \chi^{2}/dof =
615.9/502$. We note that varying $r_{cor}$ did not improve the fit and
kept this parameter fixed.  We noticed a narrow iron line near $6\kev$
in the residuals. To be able to model the narrow iron line and the
associated reflection, we used the {\tt optxagnf} model
    to describe the soft excess only by fixing $f_{pl}$ to zero and
    included a separate {\tt powerlaw} for the hard component. To
fit the iron line and reflection from cold matter like putative torus,
we included {\tt pexmon} model and tied its normalisation and $\Gamma$
to the normalisation and $\Gamma$ of power-law component,
respectively. We also fixed abundances to unity and inclination to
$50^{\circ}$ in the {\tt pexmon} model. We note that the
  geometry of the distant reflector is likely more complex than a disk
  and the inclination angle may not be meaningful.  This fit resulted
in $\chi^2/dof = 600.5/503$.  Since {\tt optxagnf} model is more
realistic than a simple {\tt powerlaw}, we switched back to inbuilt
Comptonization model in {\tt optxagnf} by removing the power-law
component but this time fixing the normalization of the {\tt pexmon}
at the best-fit values obtained in the earlier fit. We also tied the
photon index of the {\tt pexmon} to that in {\tt optxagnf} and varied
both $f_{pl}$ and $\Gamma$. The fit resulted in $\chi^{2}/dof =
600.5/503$. Now we varied the spin as well as $r_{cor}$
    together.  This improved the fit to $\chi^{2}/dof = 568.7/501$. The
best-fit parameters are listed in Table~\ref{pexopt_fit_par} and the
spectral data, the model and the residuals are shown in
Figure~\ref{jan10_fit_fig} (c).

\subsection{July 2007 \suzaku{} data}
For our spectral analysis, we used $0.6-50\kev$ band but the cross
calibration for PIN data was fixed at 1.16 rather than 1.18 as the
observation was performed at the XIS nominal position. First we
compared the XIS0, XIS1 and XIS3 data. We fitted an absorbed {\tt
  powerlaw} multiplied by a constant component to account for any
differences in the overall normalisation between different datasets
obtained from the three XIS instruments. We plotted the data-to-model
ratios in Figure~\ref{suzaku07_spec_fit} (a) after extrapolating the
$2-10\kev$ absorbed {\tt powerlaw} down to $0.6\kev$.  We found a
small discrepancy among these instruments below $0.8\kev$ and large
calibration uncertainties near $2\kev$. We therefore excluded the
$1.78-1.9\kev$ band in all our fits below.

For a comparison with the 2010 observation, we fitted an absorbed {\tt
  powerlaw} plus {\tt Gaussian} line model to the $2.5-10\kev$ XIS
spectra. The best-fit parameters are listed in
Table~\ref{par_pl1g}. It is clear that the source was brighter and the
iron K$\alpha$ line was broader and stronger in 2007 compared to that
in 2010 \suzaku{} observation. We address this spectral variability
after modelling the broadband spectra.

As before we used the $15-50\kev$ HXD-PIN data and fitted the XIS+PIN
band $0.6-50\kev$ with the {\tt wabs$\times$ powerlaw} model that
revealed different spectral components -- soft X-ray emission excess
below $2\kev$, a broad iron line near $6\kev$ and hard X-ray emission
excess above $10\kev$.  First we fitted the broadband ($0.6-50\kev$)
spectrum of 2007 with the neutral PCA model ({\tt
  wabs$\times$zpcfabs(1)$\times$zpcfabs(2)(zbbody+powerlaw+\\zgauss}). This
model resulted in $\chi^2/dof=803.7/542$ with $kT_{bb} = 115\pm7\ev$,
$\Gamma=2.06_{-0.03}^{+0.02}$,
$N_H=1.4_{-0.2}^{+0.3}\times10^{23}{\rm~cm^{-2}}$ and $C_f =
23.5_{-2.5}^{+2.4}\%$ for PCA(1),
$N_H=1.2_{-0.1}^{+0.2}\times10^{24}{\rm~cm^{-2}}$ and $C_f =
46.0_{-3.9}^{+3.6}\%$ for PCA(2).  As before, we then fitted the
broadband data with the three different models, the ionised PCA, the
blurred reflection and the intrinsic disk Comptonization model. The
complex ionised PCA model resulted in $\chi^{2}/dof =
  786.9/541$ (see Table~\ref{pcaref_fit_par} for the list of best-fit
  parameters).  The blurred reflection model with a single reflection
component ({\tt wabs(nthcomp+diskbb+kdblur*reflionx)}) resulted in a
statistically unacceptable fit ($\chi^{2}/dof = 774.9/541$) with some
nonphysical parameters .  We therefore used a reflection model with
radius dependent emissivity law and the ionisation parameter following
\cite{walton2010}.  We used the model {\tt
  wabs(powerlaw+kdblur*reflionx+kdblur*\\reflionx)} and tied the outer
radius of the first {\tt kdblur} component to the inner radius of the
second {\tt kdblur} component. The photon index of the illuminating
continuum in both the reflection components were tied to the photon
index of the power-law component. This model resulted in $\chi^{2}/dof
= 733.8/538$ and we noticed that the best-fit parameters are
consistent with that obtained by \cite{walton2010}. We replaced the
{\tt powerlaw} by {\tt nthcomp} in the composite model and fixed the
seed photon temperature to $30\ev$. This model {\tt
  wabs(nthcomp+kdblur*reflionx+kdblur*reflionx)} resulted in $
\chi^{2}/dof = 735.2/538$ (see
  Table~\ref{pcaref_fit_par}). The intrinsic disk model {\tt
  wabs(pexmon+optxagnf)} resulted in $\chi^{2}/dof~= 750.0/542$
  (see Table~\ref{pexopt_fit_par}).

\subsection{ The difference spectrum} 
Our spectral analysis of 2010 and 2007 \suzaku{} observations show
spectral variability. The reflection including the broad iron was
stronger in 2007 compared to that in 2010 (see Table~\ref{par_pl1g}
and \ref{pcaref_fit_par}).  To further investigate the variability of
the soft X-ray excess, we performed a joint spectral fitting of the
XIS spectra of the two observations. First we improved the
signal-to-noise of the XIS data by combining the XIS0 and XIS3 spectra
(XIS03) for each observation using {\it FTOOL} ``ADDASCASPEC" and then
grouped each of the combined spectra by a factor of 8. We
    modeled the soft X-ray excess with {\tt diskbb$+$zbbody} as both
    the components are required to desctibe the data statstically
    well.  We used a powerlaw for the hard component and multiplied
    all the components by the Galactic absorption.  In the joint fit
    of 2007 and 2010 XIS03 data, we tied all the model parameters
    except the powerlaw normalization. We varied the tied parameters
    together for both the datasets.  We used different power-law
    normalisations and varied separately for the two observations.
This resulted in a poor fit $\chi^2/dof = 649.6/357$. Next we varied
the normalisations of both the soft excess components separately for
both the datasets, the fit improved to $\chi^2=495.9/355$. This shows
that the soft excess and the power-law component both varied.  Next we
kept all the model components the same for both the observations but
used an overall multiplicative factor which we fixed at 1 for 2007
observation and varied for the 2010 observation. The fit resulted in
$\chi^2/dof = 499.2/357$ with the multiplicative factor of
$0.80\pm0.003$ for the 2010 spectrum. Thus both the soft excess and
power-law components varied by almost the same factor between the two
\suzaku{} observations.
   
To further establish the spectral variability in a model independent
way, we obtained a difference spectrum by subtracting the 2010 XIS
spectral data from the 2007 XIS spectral data. 
%{\color{red}{ We did
%    not used HXD/PIN data for difference spectrum. Since 2010 XIS03
%    spectral data was used as background in the difference
%    spectrum. So we used July 2007 response matrix for the modeling of
%    difference spectrum}}. 
We first fitted the difference spectrum
with an absorbed power-law model in the $2-10\kev$
band. The power-law model resulted in $\Gamma\sim 1.8$.  We
extrapolated this power-law to low energies that revealed strong soft
X-ray excess. Figure~\ref{diff_spec} shows the difference spectrum and
the soft excess below $2\kev$. Clearly, the soft X-ray excess varied
between the two \suzaku{} observations. Addition of a {\tt diskbb}
component for the soft excess improved the fit to $\chi^{2}/dof=
213.9/181$ with $kT_{in}=173\pm13\ev$, $\Gamma=1.82\pm0.04$.

\begin{figure*}
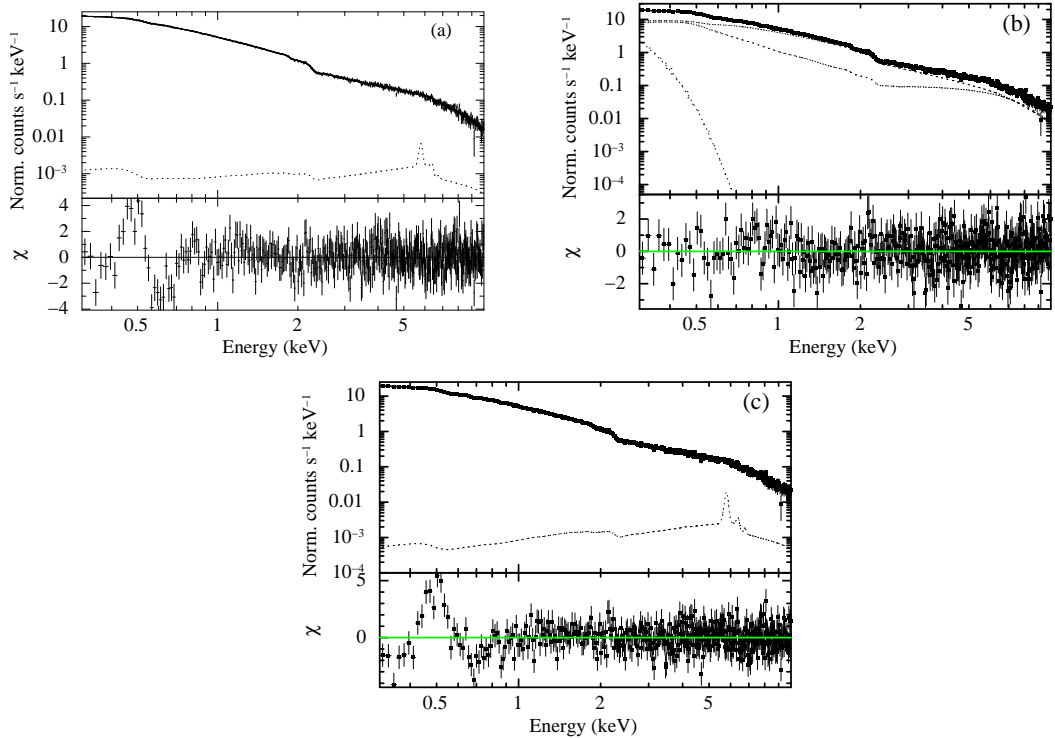

  \centering
  \includegraphics[width=5cm,angle=-90.00]{fig6a.ps}
  \includegraphics[width=5cm,angle=-90.00]{fig6b.ps}
  \includegraphics[width=5cm,angle=-90.00]{fig6c.ps}

  \caption{Results of the spectral analysis of the first \xmm{}
    observation of May 2010. The EPIC-pn spectral data, the best-fit
    model and the deviations of the observed data from the best-fit
    model -- (a) complex PCA model ({\tt wabs$\times$
          zxipcf$\times$ zxipcf(powerlaw$+$pexmon)}, (b) blurred
    reflection model ({\tt wabs(nthcomp+kdblur{*}reflionx+diskbb)})
    (c) Intrinsic disk Comptonization model ({\tt
      wabs(pexmon+optxagnf)}).}
  \label{xmm_spec_fig}
\end{figure*}
 
\subsection{\xmm{} May 2010 datasets} In addition to \suzaku{}
observations, we analysed $0.3-10\kev$ spectral data obtained from the
two observations in May 2010 as mentioned in Table~\ref{obs_log}. We
used EPIC-pn data corresponding to each observation due to high signal
to noise ratio.  We performed the spectral fitting step by step as
described above for both the \suzaku{} observations. We fitted the
three models -- the complex ionised PCA, the blurred reflection and the
intrinsic disk Comptonization model in the $0.3-10 \kev$ band. 
%
%These
%models fitted to the first observation (obs. ID: 0604720301) resulted
%in $\chi^{2}/dof=561.4/424$, $\chi^{2}/dof=460.8/422 $, and $
%\chi^{2}/dof=585.9/425$, for the ionized PCA, blurred reflection  and intrinsic disk
%Comptonization model, respectively. 
%The best-fit models for the second
%observation (Obs. ID :0604720401) resulted in $\chi^{2}/dof=521.3/420
%$, $\chi^{2}/dof=447.7/418 $, and $\chi^{2}/dof=635.0/422$, for the
%PCA, blurred reflection and intrinsic disk Comptonization model,
%respectively.  
%
These models fitted to the first observation
    (obs. ID: 0604720301) resulted in $\chi^{2}/dof=560.5/423$,
    $460.8/422$ and $585.9/425$ for the
    ionized PCA, blurred reflection ({\tt
      wabs(diskbb+nthcomp+kdblur*reflionx)}) and the intrinsic disk
    Comptonization model, respectively. The best-fit models for the
    second observation (Obs. ID :0604720401) resulted in
    $\chi^{2}/dof=585.0/419$, $447.7/418$ and
    $635.0/421$ for the ionised PCA, blurred reflection
    and the intrinsic disk Comptonization model, respectively. The
best-fit parameters are listed in Table~\ref{pcaref_fit_par} and
\ref{pexopt_fit_par}. The best-fit model, spectral data and residuals
for the first observations are shown in Figure~\ref{xmm_spec_fig}.

\section{Discussion}
We have studied the broadband X-ray spectrum of 1H~0419--577 and
confirmed the presence of the soft X-ray excess below
$2\kev$ and the hard X-ray excess above $10\kev$ earlier detected by
\cite{turner2009} and \cite{walton2010}. We also detected iron
K$\alpha$ line near $6.4\kev$. Our spectral analysis of the two
\suzaku{} and two \xmm{} observations have revealed that the spectral
components -- the soft X-ray excess, the iron line, the primary
power-law component and the hard X-ray excess all are variable.  Below
we discuss the nature of these spectral components based on their
spectral shape and variability.

\subsection{Hard X-ray excess emission}

 The neutral PCA, the ionised PCA and the reflection models provide
satisfactory fit to the two \suzaku{} observations of 2007 and
2010. In the PCA models, the observed hard excess is the result of a
Compton thick PCA of the primary powerlaw component while in the
blurred reflection model, the hard excess arises due to the reflection
hump.  We found that the reflection model resulted in statistically
better fits for both the observations (see
Table~\ref{pcaref_fit_par}).  An ad hoc blackbody component used in
the neutral PCA model for the soft excess emission resulted in a high
temperature ($kT_{BB} \sim 85\ev$ and $115\ev$ for the 2010 and 2007
observations, respectively (see below). The soft X-ray excess emission
cannot be easily explained in the framework of neutral PCA
model.  Moreover, the ionised PCA model provided poor fits to the soft
X-ray excess emission observed with \xmm{} (see
Fig.~\ref{xmm_spec_fig}).  On the other hand, the blurred reflection
model describes the hard excess and the soft excess when combined with
the tail of the thermal emission from an accretion disk. In addition,
if the hard X-ray excess emission is due to the Compton thick PCA, the
absorption corrected $2-10\kev$ luminosity of the powerlaw component
alone is $10^{45}{\rm~ergs~s^{-1}}$ for the 2010 \suzaku{} observation
and the implied bolometric luminosity is $L_{bol}\sim
10^{47}{\rm~ergs~s^{-1}}$ where we have used a bolometric correction
factor of $k_{2-10\kev}=100$ \citep{2009MNRAS.392.1124V}. This results
in $L_{bol} \sim 10 \times L_{Edd}$ for a SMBH of $M_{BH} \sim
10^{8}M_{\odot}$ estimated by \cite{pounds2004b} for 1H~0419--577.
Thus, we find that the blurred reflection is a better description both
statistically and physically for the hard X-ray excess than the
Compton thick PCA model.
    
\subsection{Nature of the soft X-ray excess emission}
Both the reflection and the intrinsic disk Comptonization models
describe the soft X-ray excess emission observed with \suzaku{}
equally well (see Table~\ref{pcaref_fit_par}). However, the intrinsic
disk Comptonization model provided poorer fit to the soft excess
emission observed with \xmm{} (see Table~\ref{pexopt_fit_par} ). The
intrinsic disk Comptonization model fitted to the \xmm{} data resulted
in an excess emission feature near $\sim 0.5\kev$, this caused the
poor quality of the fit. Since the well calibrated \suzaku{} XIS
events start at $0.6\kev$, the deviations near $0.5\kev$ are excluded
and the intrinsic disk Comptonization model provided good fits. On the
other hand, the reflection describes the hard X-ray excess emission
well, it alone cannot explain the observed soft X-ray excess
emission. Our spectral analysis has revealed that three of the four
observations require a contribution from an additional blackbody or
multicolor accretion disk blackbody with inner disk temperature
($kT_{in}\sim 20-35\ev$).  From the theory of standard accretion disks
\citep{novikov1973, sun1973}, the approximate temperature of the disk
blackbody emission is given by
\begin{equation}
  T(r)\sim 6.3\times10^{5}\left(\frac{\dot M}{\dot M_{E}}\right)^{1/4}M_{8}^{-1/4}\left(\frac{r}{R_S}\right)^{-3/4}{\rm~K}
\end{equation}
where $M_{8}$ is the mass of black hole in units of $
10^{8}M_{\odot}$, $r$ is the distance from the central black hole,
$\frac {\dot M}{\dot M_{E}}$ is the relative accretion rate.  This
equation shows that the temperature of the accretion disk can not be
beyond the extreme UV range if $r= 3R_S=6GM/c^2$ for a relative
accretion rate of 1 and a black hole mass $M_{BH} \sim
10^{8}~{M_{\odot}}$. In this case the expected inner disk temperature
is $\sim 27\ev$. For a maximally rotating black hole, the inner disk
temperature is $\sim 50\ev$. Thus, the expected disk temperature is
similar to the best-fit temperature obtained from the model with
blurred reflection.

In the reflection model, substantial fraction of the soft X-ray excess
emission is due to the blurred reflection from the partially ionised
material. However, we did not detect a broad iron line from the 2010
\suzaku{} observation.  Thus, in the framework of the blurred
reflection model, either the iron line is weaker or the disk/corona
geometry changed from 2007 to 2010 \suzaku{} observations. From
Table~2, we note that the iron line flux decreased by a factor of
about three in 2010 \suzaku{} observation compared to that in 2007
observation.

 \subsection{Variable  reflection \& the light bending model}
 1H~0419--577 varied between the two \suzaku{} observation. The soft
 ($0.6-2\kev$), hard ($2-10\kev$) and $10-50\kev$ band fluxes
 decreased by $\sim 20\%$ (see Table \ref{pcaref_fit_par}). The simple
 absorbed {\tt powerlaw} model to the $2.5-10\kev$ showed the presence
 of a broad iron line with {\tt Gaussian} $\sigma \sim 0.37\kev$ in
 the 2007 observation while such a broad line was not detected in the
 2010 observation. The iron line was weaker and narrower in January
 2010. The absence of a clear broad iron line in the 2010 data is also
 inferred from the results of best-fitting reflection model (see Table
 \ref{pcaref_fit_par}).  Overall the emissivity index is much steeper
 ($\rm q\sim 6$) and the strength of the reflection component is
 weaker in 2010 data compared to that in the 2007 data (see
 Table~\ref{pcaref_fit_par}).  The reflection spectrum inferred from
 the 2007 observation is complex and best described by a reflection
 model with radius dependent emissivity index and ionisation
 parameter. Such a complex reflection model is not statistically
 required in the case of 2010 \suzaku{} observation but the reflection
 emission arises from the innermost regions only as suggested by the
 high emissivity index of $q\sim 8$.  Thus, in the reflection model,
 the broad component of the iron line is undetected in the 2010 data
 due to extreme smearing of a weaker reflection component in the
 innermost regions of high emissivity. In the 2007 \suzaku{}
 observation, the dominant contribution to the reflection emission
 arises from the disk at intermediate radii as suggested by the second
 reflection component and relatively flatter emissivity index. This
 region is likely responsible for the increased reflection and the
 detection of broad iron line.  Thus, the iron line and the reflection
 component both weakened from 2007 to 2010 observations. The spectral
 variability is also revealed in the difference spectrum obtained by
 subtracting the 2010 spectral data from the 2007 spectral data (see
 Figure~\ref{diff_spec}). The difference spectrum shows strong soft
 X-ray excess below $2\kev$ implying that the soft excess was stronger
 in the 2007 observation. Thus, if the soft X-ray excess is
 interpreted as the blurred reflection, the observed variations imply
 variation in the reflection component as inferred from the detailed
 spectral modelling.

\begin{figure}
  \centering
  \includegraphics[width=9.2cm]{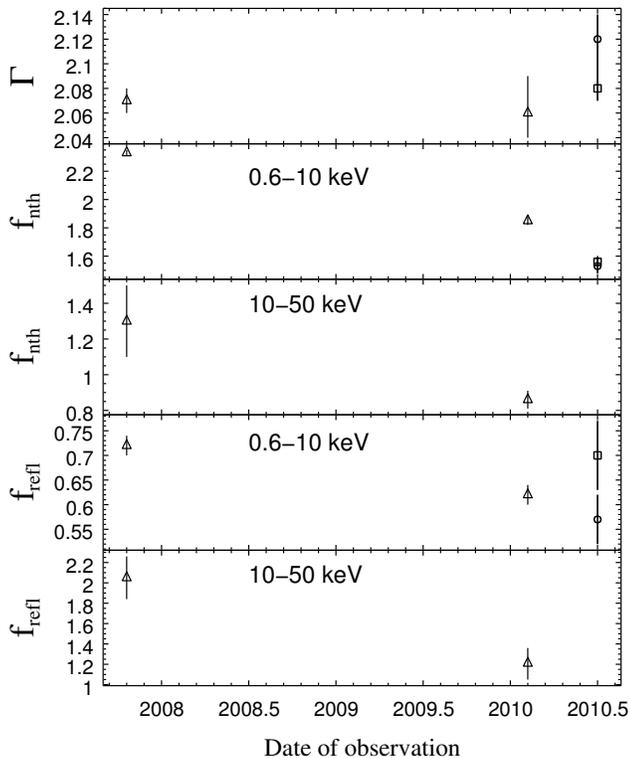}
 %\vspace{0.5cm}
  \caption{Variations in the {\tt nthcomp} and reflection emission
    based on the four observations. Triangle represent the \suzaku{}
    observations. The circle and the square represent the first and
    second \xmm{} observations of May 2010.}
  \label{spec_var}
\end{figure}

 Figure~\ref{spec_var} shows the spectral variability of 1H~0419--577.
 The reflection component appears to follow the thermal Comptonization
 component when the photon index is similar.  This is true for the
 flux in the reflection component in both the $0.6-10\kev$ and
 $10-50\kev$.  The observed correlation does not seem to hold for one
 of the \xmm{} observation when the continuum shape was slightly
 steeper. This may be due to the increased flux of the
     primary component below $\sim 2\kev$ resulting in an increased
     soft excess reflection. However, it is difficult to draw a firm
     conclusion as the spectral variability between the two \xmm{}
     observations is not very striking.

The variability properties of the reflection emission revealed in the
two \suzaku{} observations of 1H~0419--577 is different than that
observed earlier with \xmm{}.  Based on multiple \xmm{} observations
in different flux states, \citet{pounds2004a,pounds2004b} showed that
the spectral variability is dominated by a steep power-law
component. \citet{fabian2005} reanalysed the six \xmm{} observations
and demonstrated that the observed spectral variability is well
modeled by a strongly variable power-law component ($\Gamma \simeq
2.2$) and a relatively constant reflection from an ionised accretion
disk. \citet{fabian2005} identified the soft excess with the
reflection component and its weak variability resulted in the lack of
soft excess in the difference spectrum. The strongly variable
power-law and the weakly variable or almost constant reflection
component in the multiple \xmm{} observations are fully consistent
with the predictions of the light bending model \citep{fabian2005}. In
this model, the observed spectral variability is caused by the
variations in gravitation light bending depending on the changes in
the location of the power-law source with respect to the black
hole. As the power-law source moves closer to the black hole, light
bending increases and therefore the observed power-law flux drops and
the spectrum becomes more and more reflection dominated.

The spectral variability of 1H~0419--577 observed with \suzaku{} is
unusual. In the low flux state of 2010, the soft excess is weaker than
that in the high flux state of 2007. Also the iron line is weaker in
the 2010 observation. This implies that the blurred reflection became
weaker in the low flux state. Though the variability of both the soft
excess and the broad iron line seem to be consistent with the disk
reflection models where the disk responds to the continuum flux but
the weaker reflection in the low flux state is not consistent with the
variability predicted by the light bending model alone. If the
decrease in the power-law flux is solely due to the movement of the
power-law source towards the black hole, thus light bending causing
both the decrease in the observed power-law flux and steeper
emissivity index as observed. However, this should result in the
increased reflection emission which is contrary to the observations.

In the framework of reflection model, the observed spectral
variability of 1H~0419--577 with \suzaku{} can possibly be explained
in terms of changing disk-corona geometry. During the 2007 \suzaku{}
observation, the corona was likely more extended illuminating both the
inner and outer regions. In the inner regions, light bending causes
very steep emissivity index. In the regions of intermediate radii, the
light bending is less effective likely due to the extended corona and
the emissivity index is relatively flatter. Thus, the extended corona
both provides higher flux and higher reflection with relatively flat
emissivity law. During the 2010 \suzaku{} observation, most likely the
corona is more compact without the extended part resulting in
reduction of the observed power-law flux and no strong illumination to
the intermediate and outer regions of the disk. A weaker and smaller
corona results in very steep emissivity law and reduced reflection
from the inner regions only. Thus, the dynamical nature of the corona
can qualitatively explain the spectral variability of 1H~0419-577.  We
speculate that a fraction of the gravitational energy release is fed
to the hot corona in the innermost regions only if the accretion rate
is low. At higher accretion rates, gravational energy release is also
fed to larger radii, thus making the corona extended. Such a scenario
would explain both the stronger continuum and reflection including the
soft excess and the broad iron line in a high flux state.

\subsection{Spectral Variability and the intrinsic disk Comptonization model}
The intrinsic disk Comptonization model \citep[{\tt
  optxagnf}:][]{done2011} provides slightly poorer fit to both the
\suzaku{} observations compared to the blurred reflection model. Both
the models, however, require intrinsic disk emission at softest X-ray
energies. In the {\tt optxagnf} model, only the outer regions of a
disk emit thermal emission and the gravitational energy release is
split into two parts in the inner regions. A fraction of the
gravitational energy release powers the inner optically thick
Comptonised disk producing the soft excess and the other fraction
powers the hot corona. In this model, the strength of thermal outer
disk emission, the soft X-ray excess and the hard power-law depend on
the accretion rate.  Thus, in the framework of the intrinsic disk
Comptonization model, the weaker soft excess and the power-law
components in the 2010 \suzaku{} observation suggest lower accretion
rate. The absence of broad iron line in the 2010 data may be due to
the weaker illuminating power-law and/or a truncated accretion disk
possibly resulting from lower accretion rate..

 We note that the {\tt optxagnf} model provided high
  black hole spin $a>0.9$ for the two \suzaku{} observations. This
  suggests that the accretion disk extends very close to the black
  hole. Thus we expect strong blurred reflection from the disk unless
  the corona is moving away from the disk with large velocity and the 
  iron abundance is very low. However, we have shown that the blurred 
  reflection model alone best describes the observed data. This implies 
  that the dominant mechanism causing the soft excess and the spectral 
  curvature is most likely the blurred reflection.

\section{Conclusions}
We have performed detailed spectral modelling of the broadband spectra
of 1H~0419--577 obtained from two \suzaku{} observations separated by
nearly three years and two \xmm{} observations in May 2010.  We tested
three different physical models - ($i$) the complex PCA model, ($ii$)
the blurred reflection model, and ($iii$) the intrinsic disk emission.
The main results of our paper are as follows.
\begin{enumerate}%[($i$)]
\item The blurred reflection model provides statistically the best fit
  among the three models. The PCA model
  resulting in the worst statistics is unlikely to be a correct
  physical model as it results in $L_{bol} \sim 10L_{Edd}$ for
  1H~0419--577.
\item Irrespective of the models, the strong soft excess requires
  intrinsic thermal emission from an accretion disk.  Reflection
  emission alone is not sufficient to explain the observed soft
  excess.
\item We find remarkable spectral variability between the two
  \suzaku{} observation. The soft X-ray excess, the iron line and the
  power-law continuum became weaker in the 2010 compared to that
  observed in 2007. A moderately broad iron line observed in the high
  flux state of 2007 appears to have weakened in the 2010 observation.
\item Variations in the soft X-ray excess and the iron line from
  1H~0419--577 are unusual among AGNs. Such spectral variability
  demonstrates possible physical association between the soft X-ray
  excess and the iron line and is expected in the blurred reflection
  model. However, such variability is not easily explained by simple
  light bending model. Changes in the accretion disk/corona geometry
  are likely responsible for the observed spectral variability.
\item The intrinsic disk Comptonization model of Done et al. 2012 can
  also describe the broadband \suzaku{} spectra and the spectral
  variability. However, a contribution from blurred reflection is
  required. The EPIC-pn spectra below $1\kev$ are not well described
  by the intrinsic disk Comptonization model.
\end{enumerate}

We thank an anonymous referee for useful comments and suggestions. We
also thank A. C. Fabian for discussion on the complex blurred
reflection model.  This research has made use of archival data of
\xmm{} and \suzaku{} observatories through the High Energy
Astrophysics Science Archive Research Center Online Service, provided
by the NASA Goddard Space Flight Center.

% \section*{}

\newcommand{\pasp}{PASP} \def\apj{ApJ} \def\mnras{MNRAS}
\def\aap{A\&A} \def\apjl{ApJ} \def\aj{aj} \def\physrep{PhR}
\def\pre{PhRvE} \def\apjs{ApJS} \def\pasa{PASA} \def\pasj{PASJ}
\def\nat{Nat} \def\ssr{SSRv} \def\aapr{AAPR} \def\araa{ARAA}
\bibliographystyle{mn2e} \bibliography{refs}

\end{document}